\documentclass[aps,nofootinbib,groupedaddress]{revtex4}

\usepackage{url}
\usepackage{graphicx}
\usepackage[tablename=Table, figurename = Figure, font=small]{caption}
\usepackage{amsmath,amssymb,amstext,amssymb,amsfonts,amsthm}
\usepackage{hyperref}
\usepackage{rotating}
\usepackage[margin=1.0in,papersize={8.5in,11in}]{geometry}
\usepackage[utf8]{inputenc}
\usepackage{sidecap}
\usepackage{gensymb}
\usepackage[usenames, dvipsnames]{color}
\usepackage{chngcntr}
\usepackage{array,multirow}
\newcolumntype{C}[1]{>{\centering\arraybackslash}m{#1}}

\usepackage{placeins}

\def\be{\begin{equation}}
\def\ee{\end{equation}}
\def\ba{\begin{eqnarray}}
\def\ea{\end{eqnarray}}

\def\k{{\mathbf k}}

\begin{document}

\title{Towards testing CMB anomalies using the kinetic and polarized Sunyaev
Zel'dovich effects}

\author{Juan I. Cayuso${}^{1, 2}$}
\email[]{jcayuso@perimeterinstitute.ca}
\author{Matthew C. Johnson${}^{1, 3}$}
\email[]{mjohnson@perimeterinstitute.ca}

\affiliation{${}^1$Perimeter Institute for Theoretical Physics, Waterloo, Ontario N2L 2Y5, Canada}
\affiliation{${}^2$Department of Physics and Astronomy, University of Waterloo, Waterloo, ON, N2L 3G1, Canada}
\affiliation{${}^3$Department of Physics and Astronomy, York University, Toronto, Ontario, M3J 1P3, Canada}

\begin{abstract}
Measurements of the Cosmic Microwave Background (CMB) temperature anisotropies on large angular scales have uncovered a number of anomalous features of marginal statistical significance, such as: a hemispherical power asymmetry, lack of power on large angular scales, and features in the power spectrum. Because the primary CMB temperature power spectrum has been measured at the cosmic variance limit, determining if these anomalies are hints of new physics as opposed to foregrounds, systematics, or simply statistical flukes, requires new observables. In this paper, we highlight the potential contribution that future measurements of the kinetic Sunyaev-Zel'dovich effect (kSZ) and the polarized Sunyaev Zel'dovich effect (pSZ) could make in determining the physical nature of several CMB anomalies. The kSZ and pSZ effects, temperature and polarization anisotropies induced by scattering from free electrons in the reionized Universe, are the dominant blackbody contribution to the CMB on small angular scales. Using the technique of SZ tomography, measurements of kSZ and pSZ effects can be combined with galaxy surveys to reconstruct the remote CMB dipole and quadrupole fields, providing a 3-dimensional probe of large scale modes inside our Hubble volume. Building on previous work, we forecast the additional constraining power that these observables might offer for a representative set of anomaly models. We find that the remote CMB dipole and quadrupole contain a similar amount of information on anomaly models as the primary CMB polarization. The information from CMB temperature, polarization, and the remote dipole and quadrupole fields is complementary, and the full set of observables can improve constraints on anomaly models by a factor of $\sim 2-4$ using next-generation CMB experiments and galaxy surveys. This could be sufficient to definitively establish the physical origin of several CMB anomalies.
\end{abstract}

\maketitle

% \tableofcontents

\section{Introduction}

Anisotropies in the Cosmic Microwave Background (CMB) are a powerful probe of early Universe physics. On large angular scales, these anisotropies encode primordial density fluctuations, which may ultimately have been produced at energy scales far beyond the reach of any terrestrial particle accelerator. Interestingly, a series of anomalous large angular scale features in the microwave sky have been reported by the WMAP and Planck~\cite{Bennett:2003ba,Tauber2010} satellite missions, offering what could be hints of physics beyond the standard model of cosmology, $\Lambda$CDM. Several notable anomalies include: a hemispherical power asymmetry, a lack of correlations on large angular scales, features in the angular power spectrum, and an alignment of multipoles; see \cite{1510.07929} for a recent review. As the statistical significance of each of these anomalies is rather modest, the most conservative position is to attribute them to statistical flukes, given the a posteriori nature of their discovery, systematics or foregrounds. Unfortunately, as a stand-alone probe, the CMB temperature has already reached the limit imposed by cosmic variance on large angular scales, so new information can only come from alternative or complementary probes of the largest scales in the Universe. 

Several observables have been identified as potential probes of physical models of the CMB anomalies, including: CMB polarization (see e.g. Refs.~\cite{PhysRevD.77.063008,Copi:2013zja,Yoho:2015bla,ODwyer:2016xov,Bunn:2016kwh,Contreras:2017zjv,Obied:2018qdr,Billi:2019vvg}), CMB lensing (see e.g.~\cite{Yoho:2013tta,Zibin:2015ccn}), the integrated Sachs-Wolfe (ISW) effect (see e.g.~\cite{Muir:2016veb,Copi:2016hhq,Foreman:2018lci}), and probes of large scale structure in the late Universe (see e.g.~\cite{Chen:2016vvw,Chen:2016zuu,Ballardini:2017qwq,Palma:2017wxu}). Each of these observables has both advantages and disadvantages. CMB polarization can access scales comparable to those in the CMB temperature. On the largest angular scales, however, the mapping between the observed polarization anisotropies and physical scales is dependent on the (relatively poorly constrained) history of reionization. In addition, on large scales galactic foregrounds are challenging (though not impossible) to remove~\cite{Aghanim:2018eyx}. The lensing potential can be reconstructed with high fidelity using future CMB datasets (e.g. Simons Observatory~\cite{Ade:2018sbj} or CMB-S4~\cite{1610.02743}), however there is limited support from the physical scales associated with many of the CMB anomalies (see~\cite{Zibin:2015ccn}). If the (late time) ISW contribution to the CMB temperature can be isolated (e.g. using the technique of~\cite{Foreman:2018lci}), this could contribute a modest number of modes probing large scales. Finally, future galaxy surveys (e.g. LSST~\cite{0912.0201}, Euclid~\cite{Laureijs2011}, Spherex~\cite{Dore:2014cca}) or 21cm experiments (e.g. CHIME~\cite{Shaw:2013wza}, HIRAX~\cite{Newburgh:2016mwi}; see also~\cite{Ansari:2018ury}) can reach large enough volumes to offer new information on some of the CMB anomalies. While promising the measurement of a huge number of modes on linear scales, there will be limited support on physical scales responsible for the lowest multipoles of the CMB temperature, and measurement of the largest modes will be noisy and plagued by various systematics (see e.g.~\cite{Elsner:2015aga}).

The goal of this paper is to explore a new set of observables that may shed light on the physical nature of CMB anomalies: the remote dipole and quadrupole fields, i.e. the $\ell = 1,2$ moments of the microwave sky measured throughout our observable Universe. The remote dipole field can be reconstructed using the technique of kinetic Sunyaev Zel'dovich (kSZ) tomography~\cite{Ho09,Shao11b, Zhang11b, Zhang01,Munshi:2015anr,2016PhRvD..93h2002S,Ferraro:2016ymw,Hill:2016dta,Zhang10d,Zhang:2015uta,Terrana2016,Deutsch:2017ybc,Smith:2018bpn,Munchmeyer:2018eey,Sehgal:2019nmk}; the remote quadrupole field can be reconstructed using the technique of polarized Sunyaev Zel'dovich (pSZ) tomography~\cite{Kamionkowski1997,Bunn2006,Portsmouth2004,2012PhRvD..85l3540A,Hall2014,Deutsch:2017cja,Deutsch:2017ybc,Louis:2017hoh,Meyers:2017rtf,Deutsch:2018umo}. Below, we refer to the two cases more generally as SZ tomography. The kSZ effect~\cite{SZ80}, temperature anisotropies induced by scattering in the presence of a local CMB dipole, is the dominant blackbody contribution to the CMB temperature on angular scales corresponding to multipoles $\ell \gtrsim 4000$. In the presence of a local CMB quadrupole, the scattered photons are endowed with a polarization. The polarized component of the CMB arising after reionization, primarily from collapsed structures, is known as the pSZ effect (as opposed to the component sourced near decoupling and at reionization, which is simply CMB polarization). At higher order in velocity, there are a number of frequency-dependent contributions to kSZ and pSZ; these can provide additional information on large-scale inhomogeneities~\cite{Yasini:2016pby}, although we do not study them in detail here.

SZ tomography provides a tomographic reconstruction of the remote dipole and quadrupole fields by using the statistical anisotropy of the correlation between a tracer of LSS (e.g. a galaxy redshift survey) and the small-angular scale CMB temperature and polarization anisotropies. A set of quadratic estimators for the remote dipole and quadrupole fields were derived in Refs.~\cite{Deutsch:2017ybc,2012PhRvD..85l3540A,Smith:2018bpn}, and a series of forecasts, including a demonstration with simulations in Ref.~\cite{Cayuso:2018lhv}, has established detectability with future datasets~\cite{Terrana2016,Deutsch:2017cja,Deutsch:2017ybc,Smith:2018bpn} and highlighted several interesting applications including improved constraints on: primordial non-gaussianity~\cite{Munchmeyer:2018eey}, primordial gravitational waves~\cite{2012PhRvD..85l3540A,Deutsch:2018umo}, and pre-inflationary relics~\cite{Zhang:2015uta}. A key feature of SZ tomography is that it reconstructs large-scale inhomogeneities from anisotropies on the smallest angular scales. The fidelity of the reconstruction improves with the sensitivity and resolution of the CMB experiment and the depth and redshift errors of the galaxy survey. Therefore, the information accessible using SZ tomography will improve greatly with time, while direct probes of the largest scales are already close to the cosmic variance limit.

The coarse-grained remote dipole and quadrupole fields that will become accessible to the next generation of CMB and galaxy surveys are primarily sensitive to inhomogeneities on large physical scales. The remote quadrupole field receives support from the same scales contributing to the low-$\ell$ moments of the CMB temperature. Although at low redshift and on large angular scales the remote quadrupole field is strongly correlated with the primary CMB temperature quadrupole~\cite{Kamionkowski1997,Bunn2006}, there is significant new information on moderate angular scales and at high redshift~\cite{Deutsch:2017cja,Deutsch:2018umo,Deutsch:2017ybc,Louis:2017hoh}. The remote dipole field is dominated by the coarse-grained line-of-sight peculiar velocity field, and is therefore sensitive to somewhat smaller scales than the remote quadrupole. However, it can be reconstructed at far higher signal to noise, and carries a significant amount of information on scales relevant to a variety of CMB anomalies. In particular, the remote dipole field can be used to infer the density field on the largest scales at a far higher signal to noise than a direct measurement from the galaxy survey itself~\cite{Smith:2018bpn}. This implies that constraints on models of the CMB anomalies from the remote dipole field will be stronger than those from any tracer of LSS.

Here, we forecast constraints on physical models for three representative CMB anomalies: the power asymmetry, the lack of power on large angular scales, and a feature around $\ell \sim 20-30$. To quantify the utility of SZ tomography, we compare the constraining power possible with the CMB temperature and with different combinations of CMB polarization, the dipole field, and the quadrupole field. Our Fisher-based forecast includes the full covariance between observables and assumes specifications similar to next-generation CMB and galaxy surveys. We find that the remote dipole and quadrupole fields contain roughly as much additional information on the anomaly models as CMB polarization. The constraining power using all observables is a factor of 2-4 better than what can be obtained with the CMB temperature alone. Furthermore, we show that constraining power can be modestly improved by increasing the sensitivity of the CMB experiment (a $\sim 10 \%$ improvement going from a Stage-3 to a Stage-4 CMB experiment), demonstrating the potential for future improvement of constraints using SZ tomography. These improvements can in principle (assuming Gaussian likelihood functions) establish several physical models of CMB anomalies at the $\sim 5 \sigma$-level. 

The plan of the paper is as follows. In Sec.~\ref{sec1}, we review SZ tomography and describe the properties of the remote dipole and quadrupole fields. In Sec.~\ref{sec2}, we describe the details of our forecast and introduce a figure of merit which is used to quantify the potential constraining power with different combinations of observables. In Sec.~\ref{sec:results} we present the results of our forecast, and we conclude in Sec.~\ref{conclusions}.

\section{SZ tomography: reconstructing the remote dipole and quadrupole fields}\label{sec1}

SZ tomography allows us to reconstruct the line-of-sight components of the CMB dipole and quadrupole moments as observed by free electrons on our past lightcone. Here, we review the basic features of SZ tomography and the remote dipole/quadrupole fields; further details can be found in Refs.~\cite{Deutsch:2017ybc,Cayuso:2018lhv,Smith:2018bpn}. Contributions to the CMB temperature and polarization generated via the kinetic and polarized SZ effects can be expressed through the line of sight integrals

\be
\left. \frac { \Delta T } { T } \right| _ { \mathrm { kSZ } } \left( \hat { \mathbf { n } _ { e } } \right) =  \int d \chi _ { e } \ \dot{\tau} \left( \hat { \mathbf { n } } _ { e } , \chi _ { e } \right) v _ { \mathrm { eff } } \left( \hat { \mathbf { n } } _ { e } , \chi _ { e } \right), \ \ \ \ v _ { \mathrm { eff } } \left( \hat { \mathbf { n } } _ { e } , \chi _ { e } \right) \equiv \sum_{m=-1}^{1} \Theta_{1}^m \left( \hat { \mathbf { n } } _ { e } , \chi _ { e } \right) Y_{1m} \left(\hat { \mathbf { n } } _ { e } \right)
\ee
\be
( Q \pm i U ) ^ { \mathrm { pSZ } } \left( \hat { \mathbf { n } } _ { e } \right) = \frac { \sqrt { 6 } } { 10 } \int d \chi _ { e } \ \dot{\tau} \left( \hat { \mathbf { n } } _ { e } , \chi _ { e } \right) q _ { \mathrm { eff } } ^ { \pm } \left( \hat { \mathbf { n } } _ { e } , \chi _ { e } \right), \ \ \ \ q _ { \mathrm { eff } }^{\pm} \left( \hat { \mathbf { n } } _ { e } , \chi _ { e } \right) \equiv \sum_{m=-2}^{2} \Theta_{2}^m \left( \hat { \mathbf { n } } _ { e } , \chi _ { e } \right)  {}_{\mp 2}Y_{2m} \left(\hat { \mathbf { n } } _ { e } \right)
\ee
where $\hat { \mathbf { n } } _ { e }$ denotes the line of sight direction, $\chi_e$ the comoving distance to the scatterer, $\Theta_{\ell}^{m} \left( \hat { \mathbf { n } } _ { e } , \chi _ { e } \right)$ are the moments of the CMB temperature at the scatterer, and $\dot{\tau} \left( \hat { \mathbf { n } } _ { e } , \chi _ { e } \right)$ is the differential optical depth defined as
\begin{equation}
\dot{\tau} \left( \hat { \mathbf { n } } _ { e } , \chi _ { e } \right) \equiv - \sigma_T a(\chi_e) \bar{n}_e (\chi_e) \left[ 1 + \delta_e \left( \hat { \mathbf { n } } _ { e } , \chi _ { e } \right) \right]
\end{equation}
with $a _ { e } \left( \chi _ { e } \right)$ the scale factor, $\sigma_T$ the Thompson cross-section and $\delta_e \left( \hat { \mathbf { n } } _ { e } , \chi _ { e } \right)$ the perturbations about the average electron number density $\bar{n}_e(\chi_e)$. 

\begin{figure}[htb]
  \centering
     \includegraphics[width=0.7\textwidth]{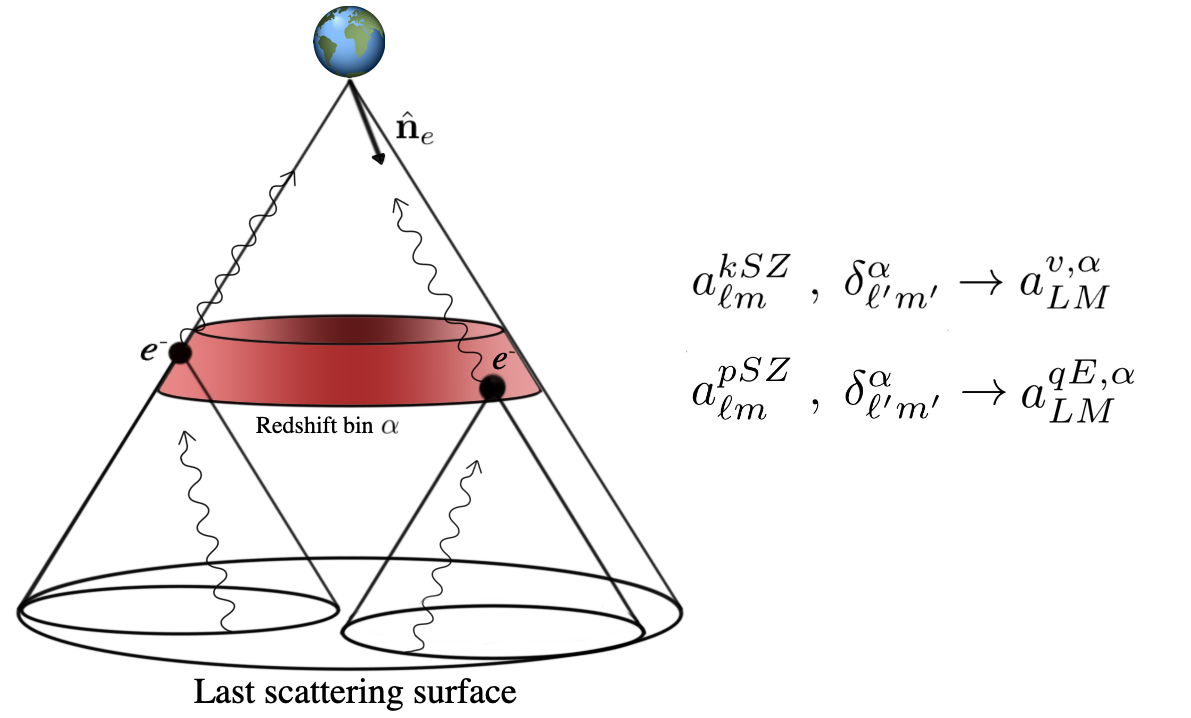}
  \caption{\label{fig:lightcone} Photons travelling from the last scattering surface can be re-scattered by free electrons once the Universe is reionized. The small scale CMB signal 
  generated through this process can be combined with a redshift dependent tracer of the electron density to reconstruct the moments $a^{v,\alpha}_{LM}$ and $a^{qE,\alpha}_{LM}$
  of the dipole and quadrupole field.}  
\end{figure}

Figure~\ref{fig:lightcone} depicts the basic spacetime geometry of the SZ effect. The remote dipole field $v _ { \mathrm { eff } } \left( \hat { \mathbf { n } } _ { e } , \chi _ { e } \right)$ is a projection of the CMB dipole $\Theta_{1}^m \left( \hat { \mathbf { n } } _ { e } , \chi _ { e } \right)$ as observed along the past light cone. The dominant contribution is from the line-of-sight component of the peculiar velocity field (as it is for our own observed CMB dipole), although there are subdominant contributions that come from the Sachs-Wolfe (SW), Integrated Sachs-Wolfe (ISW), and primordial Doppler (velocities of the plasma at last-scattering) effects. These dominant and subdominant contributions are often referred in the literature as the kinematic and intrinsic CMB dipole respectively. The remote quadrupole field $q _ { \mathrm { eff } }^{\pm} \left( \hat { \mathbf { n } } _ { e } , \chi _ { e } \right)$ is a projection of the CMB quadrupole $ \Theta_{2}^m \left( \hat { \mathbf { n } } _ { e } , \chi _ { e } \right)$ as observed along the past light cone. The remote quadrupole receives contributions from both scalar and tensor fluctuations, although we consider only scalar modes in the present context. In this case, $q _ { \mathrm { eff } }^{+} = q _ { \mathrm { eff } }^{-} $, and the remote quadrupole is curl-free. As such, we will denote the pure scalar remote quadrupole field as ``qE". The remote quadrupole is sourced by the SW, ISW, and primordial Doppler effects. For further description of the contributions to the remote dipole and quadrupole fields, we refer the reader to Refs.~\cite{Zhang:2015uta,Terrana2016,Deutsch:2017cja,Deutsch:2018umo,Deutsch:2017ybc, Seto:2005de}.

SZ tomography works by first inferring the fluctuations in the optical depth in a set of redshift bins labeled by $\alpha$ from a tracer of structure such as a galaxy survey. A quadratic estimator for the bin-averaged dipole and quadrupole fields is then constructed from the CMB temperature or polarization and each redshift bin of the galaxy survey. We work in harmonic space for the reconstructed fields, denoting the moments of the dipole or quadrupole fields in each bin as $a_{LM}^{v,\alpha}$ and $a_{LM}^{qE,\alpha}$ respectively. The reconstruction noise on the remote dipole and quadrupole fields depends on the specifications of the CMB experiment and the volume and shot noise of the galaxy survey. We discuss our assumptions for the reconstruction noise in detail in the following section, which correspond to the choices made in Ref.~\cite{Deutsch:2017ybc}. An additional consideration is the so-called ``optical depth degeneracy" (see e.g.~\cite{Hall2014,Battaglia:2016xbi}), which is a consequence of the necessarily imperfect inference of the fluctuations in the optical depth from the galaxy survey. This manifests itself as an overall multiplicative bias on the remote dipole and quadrupole fields in each redshift bin that must be marginalized over~\cite{Smith:2018bpn}. Direct measurements of the distribution of free electrons, for example using fast radio bursts as proposed in Ref.~\cite{Madhavacheril:2019buy}, can mitigate the optical depth degeneracy.

The remote dipole and quadrupole fields provide new information about the Universe on large scales. The primary CMB photons, travelling to us directly from the last scattering surface, probe the largest accessible scales. The information they provide, however, is somewhat obscured due to the fact that we observe the projection of 3-dimensional inhomogeneities onto a 2-dimensional surface. As illustrated in Fig.~\ref{fig:lightcone}, the remote dipole and quadrupole fields accessed through SZ tomography provide additional information in a number of ways. First, due to the tomographic nature of the reconstruction, we obtain coarse-grained three-dimensional information. Furthermore, the remote dipole and quadrupole fields are sensitive to inhomogeneities {\em inside our past light cone}, implying that they can access {\em different} information than what is encoded in the primary CMB temperature. In the case of the remote dipole field, which is dominated by the local peculiar velocity, it is possible to study bulk motion on scales comparable to the size of the observable Universe using long-range correlations.   

To go beyond these qualitative remarks, we inspect the scales probed by the remote dipole and quadrupole fields using linear theory, which is a good approximation for the scales under consideration. The various observables under consideration can be related to primordial gravitational potential in Newtonian gauge $\Psi_i(\boldsymbol{k})$ using a set of (bin-averaged) transfer functions $\Delta^{X,\alpha}_{\ell}(k)$ 
\be
a_{\ell m}^{X,\alpha} = \int\frac{d^3\boldsymbol{k}}{(2\pi)^3}\Delta^{X,\alpha}_{\ell}(k)\Psi_i(\boldsymbol{k})Y^{*}_{lm}(\hat{\boldsymbol{k}}),
\ee
where $X = T, E, v, qE$, corresponding to the primary CMB temperature, E-mode polarization, remote dipole, and remote quadrupole, respectively; for $X=T,E$ the index $\alpha$ is superfluous. Expressions for the remote dipole and quadrupole transfer functions, which capture  the contributions coming from the SW, ISW and Doppler effects, can be found in \cite{1610.06919,Deutsch_2018}. 

\begin{figure}[htb]
  \centering
     \includegraphics[width=0.9\textwidth]{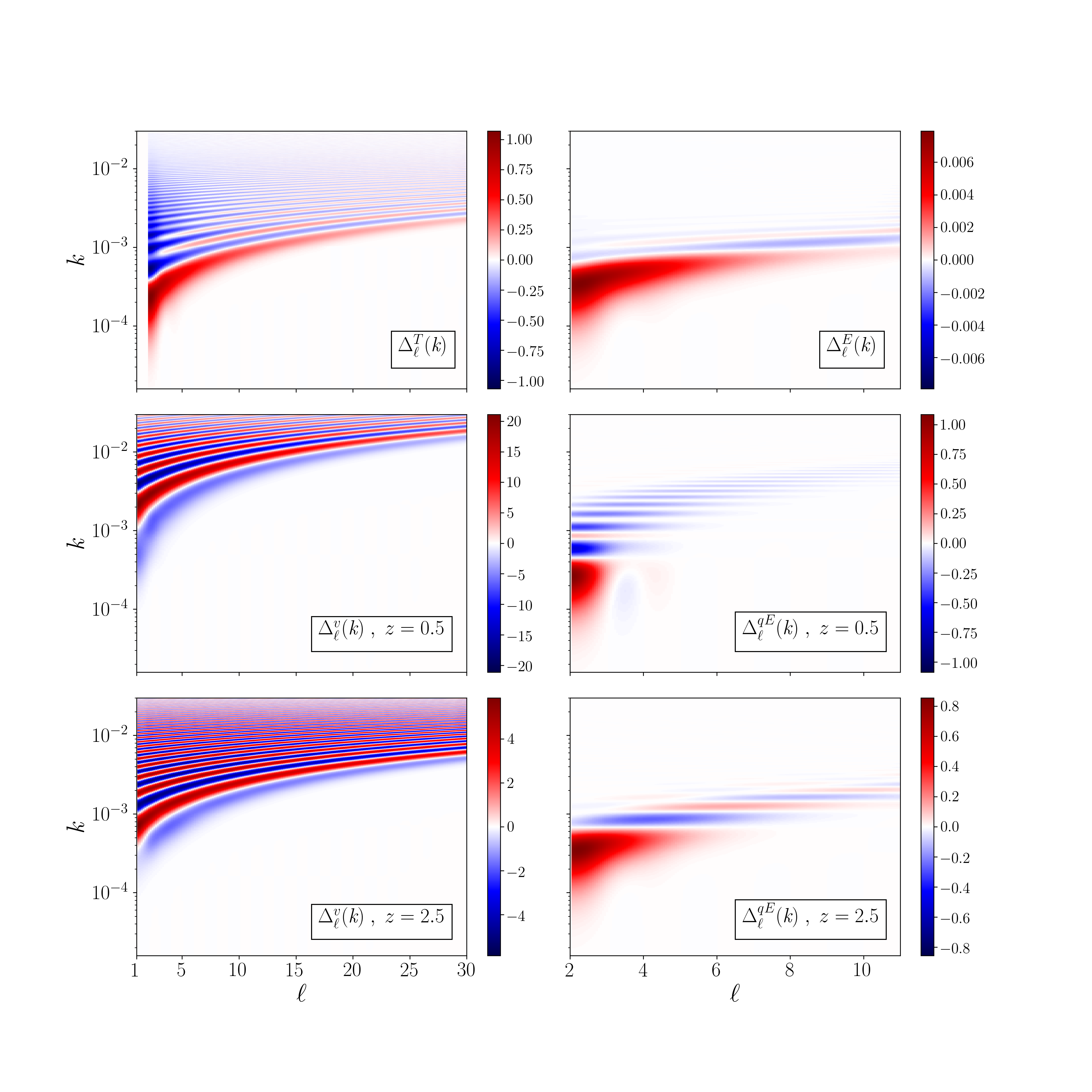}
  \caption{\label{fig:transfers}On the top panels, the transfer functions for the primary CMB temperature ($\ell =1$ is not plotted here) and E-mode polarization. On the middle and bottom, the bin averaged transfer functions for the remote dipole (left) and quadrupole (right) for bins centered on redshifts $z=0.1$ and $z=2.5$. The binning scheme 
used for this figure consisted on 60 bins of equal comoving size between $0.1 \leq z \leq 6$.}  
\end{figure}

Figure \ref{fig:transfers} shows the $\Lambda$CDM transfer functions (e.g. using parameters from Planck 2018~\cite{Aghanim:2018eyx}) in the $(\ell,k)$ plane for the primary CMB temperature, E-mode polarization and the remote fields at a few different redshifts. For the CMB temperature and remote dipole field, we plot the range $1 \leq \ell \leq 30$, which roughly encompasses the range of scales relevant to the CMB anomalies we consider. For the CMB E-mode polarization and remote quadrupole, we restrict the range to $1 \leq \ell \leq 10$, as this is the range over which the remote quadrupole receives significant support. There are a few things to note from this figure. Comparing with the CMB temperature transfer function, we see that the remote dipole and quadrupole fields have good support over a comparable range of wavenumbers. Because it is sourced mostly by fluctuations near the time of last scattering, the remote quadrupole is relatively more sensitive to large scales than the remote dipole. However, the amplitude of the remote quadrupole falls sharply with $\ell$, implying (correctly) that there will be a limited number of measurable modes. It can also be noted that the remote dipole field probes larger scales at higher redshift; this is due to the larger physical distances in the peculiar velocity field which are sampled. 

Based on this discussion, the observables returned by SZ tomography appear to have the potential to add statistical power into the analysis of CMB anomalies due to their sensitivity to large scale inhomogeneities. However, the amount of \textit{new} information that can be added will depend on the correlations that exist amongst all the observables we consider. Indeed, some correlations are expected to be there by construction. For example, at low redshift the $\ell =2$ moments of the remote quadrupole field are perfectly correlated with the CMB temperature quadrupole~\cite{Kamionkowski1997,Bunn2006,Deutsch:2017cja}. 

\begin{figure}[htb]
\centering
\includegraphics[width=0.8\textwidth]{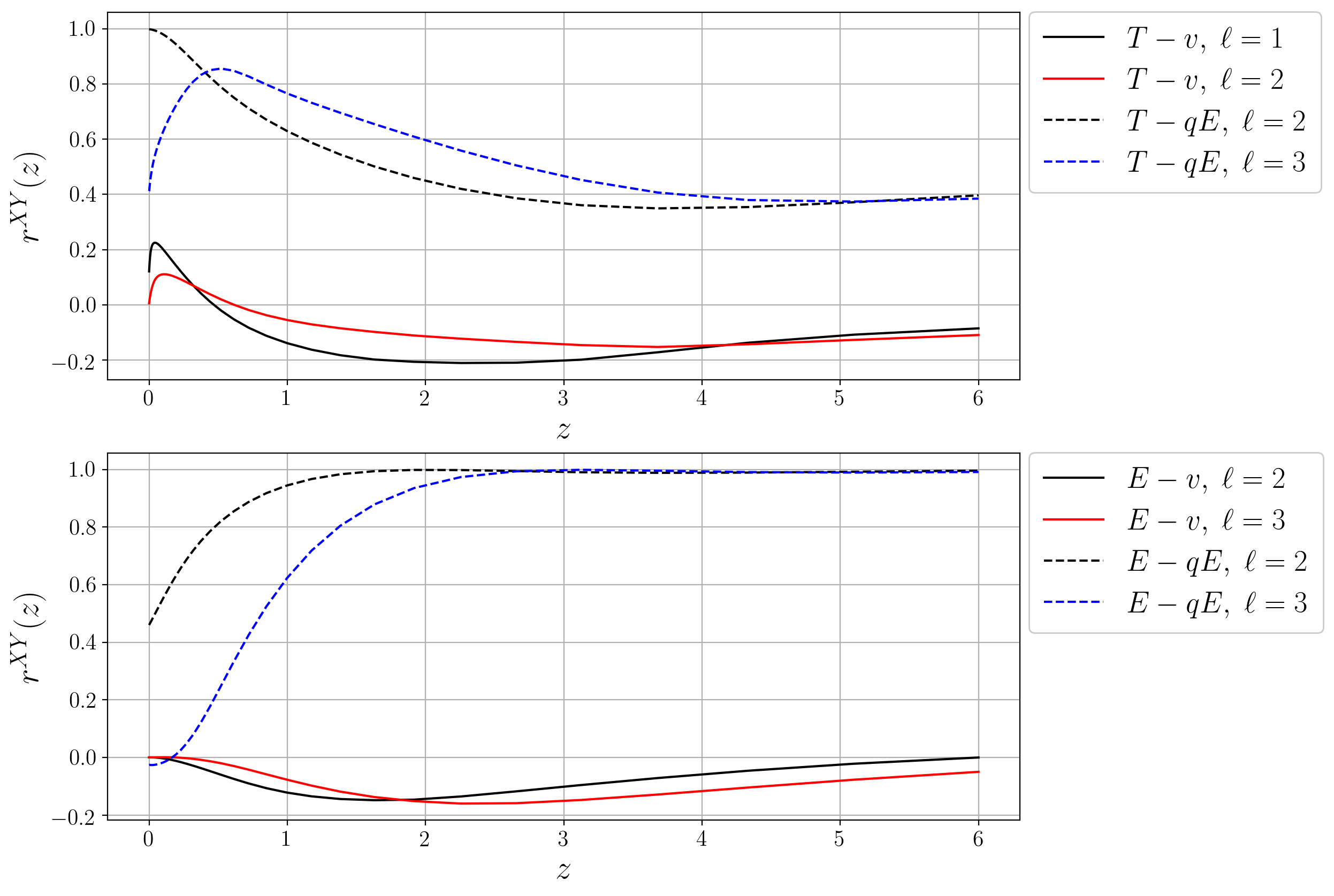}
\caption{\label{fig:correlations} Correlation coefficient between the primary CMB fields and the remote dipole an quadrupole fields. As expected, the $\ell = 2$ moment of the primary CMB temperature is perfectly correlated with the very low redshift remote quadrupole (top panel, black dashed line) and the remote dipole captures the primordial contributions to the $\ell =1$ aberration-free CMB dipole measured at $z=0$ (top panel, black solid line). The remote quadrupole exhibits longer range correlations with the primary CMB than the remote dipole does (bottom panel). } 
\end{figure}

In Figure~\ref{fig:correlations} we plot the correlation coefficient between the remote fields and the primary CMB, defined by 
\be
r_{\alpha\beta,\ell\ell',mm'}^{X,Y} = \frac{C_{\alpha\beta,\ell\ell',mm'}^{X,Y}}{\sqrt{C_{\alpha\alpha,\ell\ell,mm}^{X,X}C_{\beta\beta,\ell'\ell',m'm'}^{Y,Y}}},
\ee
where 
\be\label{covariance_elements}
C_{\alpha\beta,\ell\ell',mm'}^{X,Y} = \int\frac{d^3\boldsymbol{k}}{(2\pi)^3}\int\frac{d^3\boldsymbol{k'}}{(2\pi)^3}\;\Delta^{*X,\alpha}_{\ell}(k)\;\Delta^{Y,\beta}_{\ell'}(k')\big{<}\Psi^{*}_i(\vec{\boldsymbol{k}})\Psi_i(\vec{\boldsymbol{k'}})\big{>}
Y_{\ell m}(\hat{\boldsymbol{k}})Y^{*}_{\ell'm'}(\hat{\boldsymbol{k}}')
\ee
are the elements of the covariance matrix. In the top panel, we show the correlations between the CMB temperature and remote fields at a few values of $\ell$. For $\ell = 1$, the CMB temperature is the aberration-free dipole (see e.g. Refs.~\cite{Meerburg:2017xga,Weak_lensing,PhysRevD.78.123529,Yasini:2017jqg} for a summary of the various frames for the CMB dipole), not the dipole observed in the Earth's rest frame. There is $\alt 10-20\%$ correlation between the CMB temperature and the remote dipole field over a range of redshifts and multipoles. There is a far higher correlation between the CMB temperature and the remote quadrupole field. As expected, there is nearly perfect correlation between the CMB quadrupole and the $\ell = 2$ moment of the remote quadrupole field, except at the highest redshifts. The remote dipole field has little correlation with the CMB E-mode polarization. However, at the highest redshift, there is a near-perfect correlation between the E-mode polarization and the remote quadrupole field. This is expected, since at high redshift the CMB polarization is sourced by the same remote quadrupole field that is being reconstructed by SZ tomography (see Ref.~\cite{Deutsch:2018umo} for further discussion). In conclusion, including the full covariance between the various observables can be important in a joint analysis, such as the one we present below. This is particularly important at low-multipoles/low-redshift for the CMB temperature, and at high-redshift for the CMB E-mode polarization. Conversely, we see that over a wide range of multipoles and redshifts, the remote dipole and quadrupole fields carry significant independent information beyond the primary CMB temperature and polarization.

\section{Forecast setup}\label{sec2}

The reconstruction of the the remote dipole and quadrupole fields using SZ tomography opens the possibility of bringing new information into the study of the large scale CMB anomalies. 
Determining how informative a data set is will depend, of course, on the type of questions we are trying to answer. From the Bayesian perspective, we might strive for model-selection: How would adding a new observable change the odds-ratio between the anomaly model and $\Lambda$CDM in a future experiment (see e.g.~\cite{Trotta:2007hy})? Such an approach requires a motivated set of theoretical priors, as well as an understanding of the full likelihood function over model parameters. Due to the lack of strongly motivated models and the computational complexity of evaluating the full likelihood function, we do not pursue this approach here.  Another possibility (less computationally expensive than model selection) for predicting how informative a data set can be is to determine its constraining power on the parameters of a model using a Fisher matrix-based approach. In general, such results are not sufficient to decide if a future experiment could choose among competing theoretical models. However, this approach does offer a way of quantifying the additional constraining power a new observable might add. We will adopt this methodology in order to study the information content of the remote dipole and quadrupole field on a series of models for CMB anomalies, and compare it to what is achievable using the primary CMB temperature and E-mode polarization on large angular scales. 

Given a cosmological model with parameters $\{ \lambda_i\}$, one can forecast how well these parameters can be constrained using different set of observables by implementing a Fisher matrix analysis. The elements of the Fisher matrix are given by the following expression:
\be
{\bf F}_{ij} = \frac{1}{2}\mbox{Tr}[ (\bf{C+N})^{-1} \bf{C}_{,\lambda_{i}} (\bf{C+N})^{-1} \bf{C}_{,\lambda_{j}} ]
\ee
where $\bf C$ is the signal covariance matrix (whose elements are defined by Eq.~\ref{covariance_elements}), $\bf{C}_{,\lambda_{j}}$ denotes its derivative respect to $\lambda_j$, and $\bf N$ is the noise covariance matrix. The Fisher matrix encodes information about the curvature of the likelihood function around its maximum in parameter space, and this information can be turned into fully marginalized constraints on the model parameters:
\be
\sigma_{\lambda{i}} = \sqrt{\big{(}{\bf{F}}^{-1}\big{)}_ {ii}}.
\ee

In the analysis below, we include the $\Lambda$CDM parameters and the free parameters $a_i$ in a variety of anomaly models. In addition, we need to include a bias parameter $b_{\alpha}$ multiplying the multipole coefficients  $a_{LM}^{v,\alpha}$ and $a_{LM}^{qE,\alpha}$ in each bin $\alpha$ due to the optical depth degeneracy in kSZ/pSZ : having incomplete inference of the electron-galaxy cross spectrum, we can only reconstruct the remote dipole and quadrupole inside each redshift bin up to an overall amplitude. We refer the reader to \cite{Smith:2018bpn} for a more detailed discussion of the optical depth degeneracy. Overall, our parameter set is given by 
\begin{equation}
\lambda = \{\Omega_{b},\Omega_{c},h,\tau, A_{s},n_{s}, b_{\alpha}, a_i \},
\end{equation}
The constraints on the anomaly model parameters $a_{i}$ presented below are fully marginalized over all cosmological parameters and the optical depth degeneracy. The fiducial values for the $\Lambda$CDM cosmology we take are $\Omega_bh^2 = 0.02233$, $\Omega_ch^2 = 0.1198$, $h = 0.675$, $\tau = 0.054$, $10^9 A_s = 2.096$ and $n_s = 0.9652$. . The bias parameters $b_{\alpha}$ are assigned fiducial values of unity. Priors on the bias parameters $b_{\alpha}$ can come from other astrophysical probes~\cite{Madhavacheril:2019buy}, but for the anomaly models under consideration we find that the constraints are relatively insensitive to the addition of such a prior. Fiducial values for each of the anomalies model parameters are presented together with the models in the next section and Appendix~\ref{sec:models}.

The signal covariance matrix $\bf{C}$ we construct is split into two pieces: $\bf{C}^{\mbox{\scriptsize{low}}}$ and $\bf{C}^{\mbox{\scriptsize{high}}}$. For multipoles $\ell \leq 60$, we include all observables, accounting for the correlations between the primary CMB and remote dipole/quadrupole fields in the off-diagonal entries of $\bf{C}^{\mbox{\scriptsize{low}}}$. For multipoles $60 < \ell < \ell_{\mbox{\scriptsize{high}}}$, where the reconstruction of the remote dipole and quadrupole fields is poor, and where correlations with the primary CMB are vanishingly small, we include only the CMB temperature and polarization (and their covariance) in $\bf{C}^{\mbox{\scriptsize{high}}}$. We choose 
$\ell_{\mbox{\scriptsize{high}}}=4000$ since for higher multipoles the primary CMB becomes a subdominant contribution to the measured microwave sky. With these assumptions, the Fisher matrix factorizes 
into the sum of a low-$\ell$ and high-$\ell$ piece, $\bf{F}^{\mbox{\scriptsize{low}}}$ and $\bf{F}^{\mbox{\scriptsize{high}}}$, respectively. The main effect of $\bf{F}^{\mbox{\scriptsize{high}}}$ is to constrain the $\Lambda$CDM parameters. We further assume that $\bf{F}^{\mbox{\scriptsize{high}}}$ is zero for the entries corresponding to the anomaly model parameters $a_i$, since the anomaly models under consideration will have little to no effect on these scales.

To construct the noise covariance matrix we take into consideration specifications for the instrumental noise and shot noise for next generation CMB experiments and galaxy surveys. These are central elements in computing the reconstruction noise for the remote dipole and quadrupole fields. The reconstruction noise is computed as in Ref.~\cite{Deutsch:2017ybc}. We assume CMB temperature and polarization data on the full sky with noise
\be\label{eq:cmbnoise}
{\mbox{\bf{N}}}_{\ell}^{\mbox{\scriptsize{CMB}}} = \eta^{2} \exp\Bigg{(}\frac{\theta^{2}_{\mbox{\scriptsize{FWHM}}}}{8\mbox{log}2}\ell(\ell+1)\Bigg{)}.
\ee
We choose fiducial values of $\eta = 1\mu K$-arcmin and $\theta_{\mbox{\scriptsize{FWHM}}}=1$ arcmin, representative of Stage-4 CMB experiments, and explore how constraints vary for larger ($5 \mu K$-arcmin) and smaller ($0.1 \mu K$-arcmin) noise at fixed beam size. The CMB noise enters into the calculation of the noise covariance matrix in two different ways. First, it captures the instrumental noise for the primary temperature and polarization used in the construction of $\bf{C}^{\mbox{\scriptsize{high}}}$. Second, the instrumental CMB noise is one of the necessary pieces to calculate the reconstruction noise (see Ref.~\cite{Deutsch:2017ybc}) for the dipole and quadrupole correlations appearing in $\bf{C}^{\mbox{\scriptsize{low}}}$. Multipoles up to $\ell = 10000$ (assumed to be accessible with next generation experiments) are  used to calculate this reconstruction noise, and thus different choices of the CMB noise level $\eta$ will have an impact on the signal to noise for the low-$\ell$ dipole and quadrupole fields. The other components involved in the computation of the reconstruction noise are the details of the galaxy survey and the model for the electron distribution. The number density and fiducial galaxy bias model we incorporate is as in Ref.~\cite{Deutsch:2017ybc}, based on the expected specifications of LSST~\cite{0912.0201}. We assume that electrons are an unbiased tracer of dark matter, with the uncertainty in this assumption folded into the optical depth bias which we marginalize over. We further assume that we have data from the galaxy survey and CMB experiment on the full sky.

Constructing the signal and noise covariance matrices involves a choice of redshift binning for the galaxy survey, which determines how coarse-grained the reconstructed remote dipole and quadrupole fields are. The thinner the binning is, the more information can be collected. Clearly all independent information would be captured in the limit of having infinitely small redshift bins, but the redshift resolution of the (photometric redshift) surveys used in the reconstruction process imposes a limit on how many redshift bins can be used. We use 60 redshift bins of equal comoving radial width between $Z=[0.1,6]$, which, translating to redshfit, is roughly consistent with the expected photometric redshift errors for LSST~\cite{0912.0201}. 

Although it is sufficient to present the constraints for each model parameter, we would like to define a single quantity that facilitates comparison of constraining power among
models with different number of parameters. For a forecast using a set $\{X\}$ of observables on a model with $N$ parameters, we define the figure of merit FoM by:
\be
\mbox{FoM}(X) = {\Bigg{(}\frac{1}{\prod_{i=1}^{N}\sigma_{\lambda{i}}(X) }\Bigg{)}^{\frac{1}{N}}}.
\ee
Furthermore, since we want to highlight the relative performance respect to the primary CMB, we will express our results in terms of a figure of merit ratio, defined by :
\be\label{FoMr}
\mbox{FoMr}(X) = \frac{{\mbox{FoM}}(X)}{{\mbox{FoM}}(T)}
\ee
The figure of merit ratio encodes the geometrical mean improvement on model parameter constraints. Similar figure of merit ratios have been used in previous literature, e.g. as a measure of improvement in costraints on dark energy parameters when comparing current to future missions~\cite{2009arXiv0901.0721A}. Using these definitions, we will explore the information content on CMB anomalies models of different subsets of observables: $X=(T,E)$, $X=(T, v, qE)$ and $X=(T, E, v, qE)$.

\section{Results}\label{sec:results}

In what follows we present the results of our Fisher forecast for the additional constraining power of the remote dipole and quadrupole fields on models of the large scale CMB anomalies. We consider two general classes of physical models: models that break statistical isotropy, which could be responsible for the power asymmetry, and models that deviate from a nearly scale-invariant primordial power spectrum, which could be responsible for a lack of power on large scales and a feature in the power spectrum at $\ell \sim 20-30$. While this is a small subset of physical models considered to explain only a subset of the CMB anomalies, we hope that the cases we do consider are illustrative of the potential utility of SZ tomography for providing further insight into the nature of the CMB anomalies. For each model we present the figure of merit ratio FoMr given by Eq.~(\ref{FoMr}), which provides a quantitative measure of the overall improvement on parameter constraints relative to what is achievable with measurements of the primary CMB temperature only. The fully marginalized parameter constraints for each model can be found in Appendix~\ref{sec:models}.

\subsection{Statistical isotropy breaking}

A subset of the observed CMB anomalies suggest the existence of statistical anisotropies~\cite{Planck2015_isotropy}: unexpected alignment between the  low multipole moments, a hemispherical power asymmetry, parity asymmetry of the CMB etc. It is still not known whether or not these features are due to foregrounds, local cosmic structure or possible statistical flukes present our observed realization of $\Lambda$CDM. However, if due to true physical departures from $\Lambda$CDM, the underlying model must break statistical isotropy. 

We consider phenomenological models of spontaneous isotropy breaking~ \cite{PhysRevD.72.103002} (see also e.g.~\cite{Erickcek:2008sm,Liddle:2013czu,Kobayashi:2015qma,Adhikari:2015yya}), in which local observers would detect statistical anisotropy, while the Universe as a whole is globally statistically homogeneous. Following Ref.~\cite{PhysRevD.77.063008}, we include a field $h(\vec{\boldsymbol{x}})$ with super horizon fluctuations that modulates the potential $g_{1}(\vec{\boldsymbol{x}})$ only on large scales, leaving small scale fluctuations described by $g_{2}(\vec{\boldsymbol{x}})$ unaffected:
\be
\Psi_i(\vec{\boldsymbol{x}}) = g_{1}(\vec{\boldsymbol{x}})(1+h(\vec{\boldsymbol{x}}))+g_{2}(\vec{\boldsymbol{x}}),
\ee
Here, $g_{1}(\vec{\boldsymbol{x}})$ and $g_{2}(\vec{\boldsymbol{x}})$ are random Gaussian fields, while $h(\vec{\boldsymbol{x}})$ is deterministic within our Hubble volume. It is the slow variation of $h(\vec{\boldsymbol{x}})$ inside our Hubble volume that is responsible for the existence of statistical anisotropy in the CMB. Such a modulation can occur, for example, in inflation models with more than one field contributing to the primordial curvature perturbations. Here, rather than focusing in a particular early Universe mechanism for generating the modulation, we are interested in determining how the imprints of a preferred direction on the remote dipole and quadrupole fields can help to constrain the amplitude of the modulation.

The effect on the primordial power spectrum is given by:
 \begin{eqnarray}\label{P_mod}
\big{<}\Psi^{*}_i(\vec{\boldsymbol{k}})\Psi_i(\vec{\boldsymbol{k'}})\big{>}  &=&  (2\pi)^{3}\delta(\vec{\boldsymbol{k}}-\vec{\boldsymbol{k'}})(P_{g_1}(k)+P_{g_2}(k') )\nonumber \\
&+&  (P_{g_{1}}(k)+P_{g_{1}}(k'))h(\vec{\boldsymbol{k}}-\vec{\boldsymbol{k'}}) \nonumber \\
&+&  \int\frac{d^3\tilde{k}}{(2\pi)^3}P_{g_{1}}(\tilde{k})h(\vec{\boldsymbol{k}}-\vec{\tilde{\boldsymbol{k}}})h(\vec{\boldsymbol{k'}}-\vec{\tilde{\boldsymbol{k}}}) ,
\end{eqnarray}
where $P_{g_1}(k)$ and $P_{g_2}(k)$ are the power spectra for $g_{1}(\vec{\boldsymbol{x}})$ and $g_{2}(\vec{\boldsymbol{x}})$. We fix these power spectra to reproduce the $\Lambda$CDM power spectrum when $h(\vec{\boldsymbol{x}})=0$. The second and third term will induce couplings between different $(\ell,m)$ multipoles and this manifests the breaking of statistical isotropy for local observers. 

In what follows we will consider a dipolar modulation given by a super-horizon scale mode varying in the direction of the $z$ axis:
\be\label{eq:modulation}
h(\vec{\boldsymbol{x}}) = A\frac{\mbox{sin}{(\vec{\boldsymbol{k}}_0\cdot\vec{\boldsymbol{x}}})}{k_0\,\chi_{dec}} \approx A\frac{z}{\chi_{dec}}
\ee
This physical model could explain the observed power asymmetry~\cite{PhysRevD.77.063008} (see also e.g.~\cite{Contreras:2017qkc,Zibin:2015ccn,Contreras:2017zjv}); we do not consider other modulation models here, which could be responsible for the observed alignment of low-$\ell$ multipoles (see e.g.~\cite{PhysRevD.77.063008}) or other observed CMB anomalies. For the  modulating field Eq.~\ref{eq:modulation}, expressions for elements of the covariance matrix up to second order in $A$ can be obtained analytically and are presented in Appendix~\ref{sec:modecoupling}. Analysis of temperature data by Planck~\cite{Planck2015_isotropy} suggests a phenomenological dipole modulation up to $\ell \sim 60$ with a value for the amplitude parameter of approximately~\cite{Planck2015_isotropy} $A = 0.07 \pm 0.02$. An open question is to what scales the asymmetry might persist, e.g. where the cross-over occurs from the observed fluctuations being sourced by $g_1$ to being sourced by $g_2$. There is a hard upper bound implied by the low hemispherical asymmetry of the distribution of high redshift quasars \cite{Hirata_2009} of $k\lesssim 1 \ \mbox{Mpc}^{-1}$. In the following we treat the cross-over to statistical homogeneity by setting $P_{g_2}(k) = 0$ for $\ell < 60$ in all observables, and setting $P_{g_1}(k) = 0$ for $\ell > 60$ (at which point we only include the primary CMB temperature and polarization in our forecast). A more physical treatment would involve introducing a new set of parameters describing the cross-over, as e.g. in Ref.~\cite{Zibin:2015ccn}. However, our prescription is not in conflict with the small-scale bound from quasars (since the contribution from Mpc scales is negligible to the remote dipole, remote quadrupole, and primary CMB on large scales), and offers a quantitative measure of the additional constraining power from SZ tomography.

Table.~\ref{fig:Dipolar} shows the forecasted figure of merit ratio FoMr for different combinations of observables and CMB noise levels. First, we can compare the additional constraining power from E-mode polarization to that offered by the remote dipole and quadrupole fields (e.g. FoMr$(T,E)$ vs FoMr$(T,v,qE)$). We see that the E-mode polarization and the remote dipole/quadrupole fields carry roughly equivalent additional constraining power. However, there is significant improvement on the constraint from the remote dipole/quadrupole fields as the noise of the CMB experiment is decreased, while the E-mode polarization is already cosmic variance limited on the relevant scales. Most of the additional constraining power offered by SZ tomography comes from the remote dipole field, which has many more measurable modes. Including all observables (e.g. FoMr(All)), it is possible to roughly double the constraining power offered by the CMB experiment alone. This corresponds to the constraint $\sigma_A= 0.01$. Depending on the shape of the fully marginalized posterior for $A$ (close to Gaussian or not), this could confirm that a cosmological model with this type of modulation is preferred to $\Lambda$CDM by more than $5\sigma$. This motivates an analysis of the full likelihood function, which could in addition be use to forecast the results of model selection with future data.

\vspace*{10pt}
\begin{table}[htb]
\centering
\resizebox{0.7\textwidth}{!}{
\begin{tabular}{|c|c|c|c|}
\hline
$\;\;\;$Noise [$\mu$-Karcmin]$\;\;\;$ & $\;\;\;$FoMr($T,E$)$\;\;\;$ & $\;\;\;$FoMr($T,v,qE$)$\;\;\;$ & $\;\;\;$FoMr(All)$\;\;\;$ \\
\hline
5.0 & 1.40 & 1.48 & 1.77 \\ 
\hline
1.0  & 1.40 & 1.80 & 2.05 \\
\hline
0.1 & 1.40 & 1.87 & 2.11 \\
\hline
\end{tabular}}
\caption{Figure of merit ratio for the dipolar modulation model. Since this is a one parameter model, the FoMr informs us directly about the improvement of the $1-\sigma$ constraints.}
\label{fig:Dipolar}
\end{table}

\subsection{Deviations from $\Lambda$CDM power law}

The other class of models we consider involve possible deviations from the $\Lambda$CDM power law primordial power spectrum. On large angular scales it has been observed by WMAP and Planck that the CMB temperature shows an unexpected lack of variance compared to $\Lambda$CDM. Features in the temperature power spectrum have also been identified, remarkable ones being a low quadrupole and a lack of power at multipoles $\ell \sim 20-30$. One simple and theoretically interesting possibility is that these CMB anomalies are due to corresponding features in the primordial power spectrum of curvature fluctuations. Such features can arise as a signature of: the onset of inflation (e.g.~\cite{Linde:1998iw,Yamauchi:2011qq,Contaldi_2003}), oscillations~\cite{Chen:2006xjb,Silverstein:2003hf} or sharp steps~\cite{Adams:2001vc} in the inflaton potential, steps in the sound speed~\cite{Achucarro:2010da} or DBI inflation warp factor~\cite{Bean:2008na}, among other scenarios. In this section, we determine the additional constraining power offered by SZ tomography for a subset of these feature models, choosing a few representative examples that have previously been investigated by Planck~\cite{Planck_inflation_2015}.

\subsubsection{Phenomenological models for large scale power suppression}

\begin{figure}[htb]
  \centering
    \includegraphics[width=0.6\textwidth]{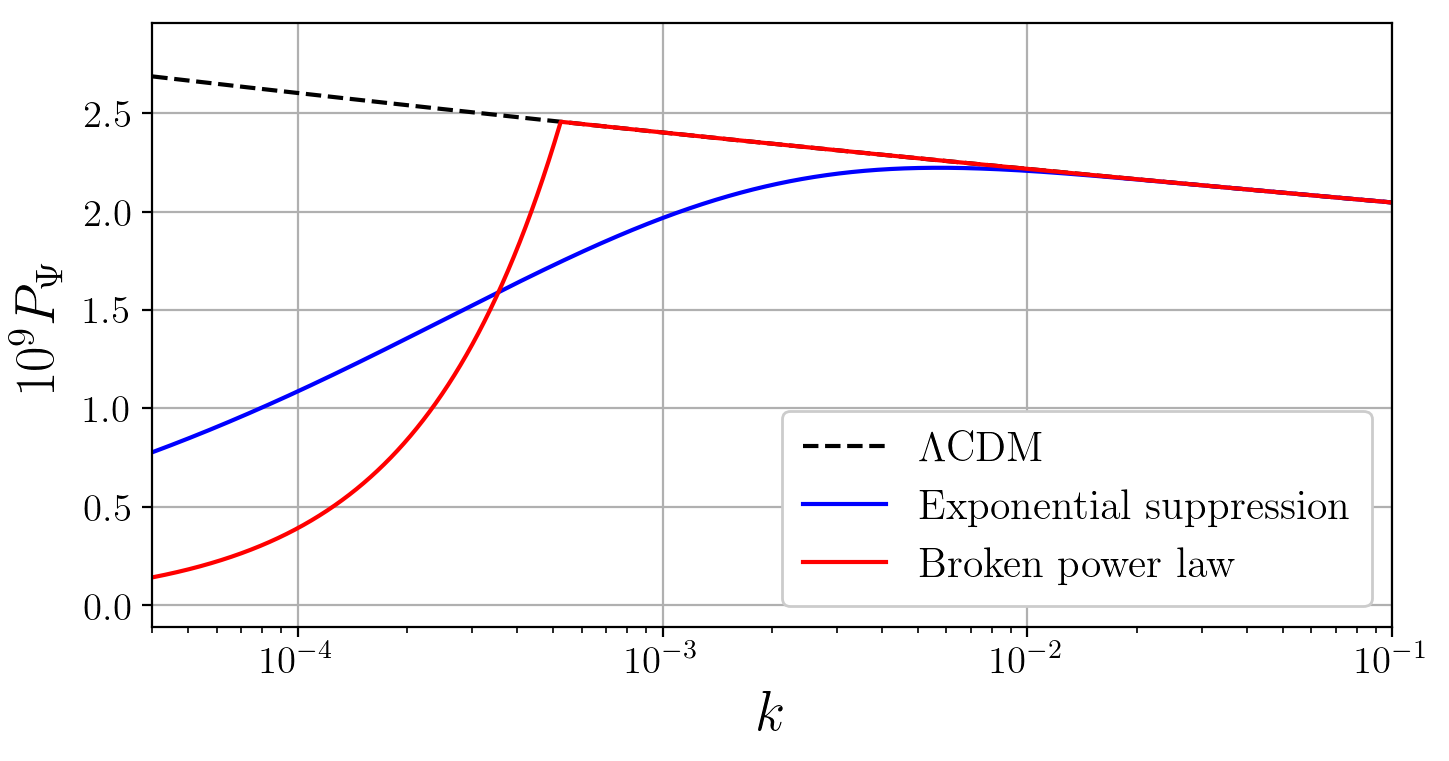}
  \caption{\label{fig:Large_sup_p} Primordial power spectrum for the exponential suppression model and the broken power law model together with the standard $\Lambda$CDM
  spectrum. The first model shows suppression starting at scales of several hundred Mpc while the second one deviates from the standard power law on scales of several Gpc.}
\end{figure}

Following Ref.~\cite{Planck_inflation_2015}, we consider a set of two-parameter phenomenological models for the suppression of power on large angular scales in the primary CMB temperature. The first model we consider~\cite{Contaldi_2003} implements an exponential suppression of power below a wavenumber $k_{c}$:
\be
P_{s_{1}}(k) = P_{0}(k)\Bigg{(}1-\mbox{exp}\bigg{[}-\bigg{(}\frac{k}{k_{c}}\bigg{)}^{\lambda}\bigg{]}\Bigg{)}
\ee
where $P_{0}(k) =  A_{s}\Big{(}\frac{k}{k_{*}}\Big{)}^{n_s-1} $ and the best-fit model parameters from Planck 2015~\cite{Planck_inflation_2015} are $k_c = 3.74 \times 10^{-4} \ {\rm Mpc}^{-1}$ and $\lambda = 0.53$. The second model has a break in the power law at a scale $k_b$:
\begin{equation}
        P_{s_{2}}(k)=
        \begin{cases}
            A_s\Big{(}\frac{k_b}{k_*}\Big{)}^{-\delta}\Big{(}\frac{k}{k_{*}}\Big{)}^{n_s-1+\delta} & \mbox{if } k\leq k_b, \\
            A_{s}\Big{(}\frac{k}{k_{*}}\Big{)}^{n_s-1} & \mbox{if } k\geq k_b, 
        \end{cases}
\end{equation}
The best-fit model parameters from Planck 2015~\cite{Planck_inflation_2015} are $k_b = 5.26 \times 10^{-4} \ {\rm Mpc}^{-1}$ and $\delta = 1.14$. In both cases, we choose the central values for the model parameters as the best-fit Planck values, we fix the pivot scale to $k_* = 0.05 \; \mbox{Mpc}^{-1}$, and (as throughout the paper) we marginalize over the $\Lambda$CDM parameters, including $A_s$ and $n_s$. We plot the two fiducial models in Fig.~\ref{fig:Large_sup_p}. Table~\ref{fig:large_sup} shows the FoMr for each model, for various combinations of datasets. In detail, we can conclude the following:
\begin{itemize}
\item \bf{Exponential suppression}: \normalfont For this model the remote fields offer slightly better improvements on constraints compared to the E-mode polarization for the CMB noise levels considered. Combining all observables, the FoMr is significant over the full range of CMB noise levels we have considered. Furthermore, there is significant improvement as the CMB noise is lowered. A detailed investigation shows that the remote dipole field drives the improvement on constraints, with the remote quadrupole contributing only marginally. At the lowest noise level considered, it is possible to measure the suppression scale $\k_c$ at the $\sim 55\%$-level and the steepness parameter $\lambda$ at the $\sim 27\%$-level. There is a significant degeneracy between the scale and steepness parameters. 

\item \bf{Broken power law}: \normalfont Unlike the exponential suppression model, deviations from $\Lambda$CDM for the broken power law model arise on scales where the remote dipole is less sensitive than E-mode polarization ($k \lesssim 5\times10^{-4}\mbox{Mpc}^{-1}$ ); see Fig.~\ref{fig:transfers}. The remote quadrupole field is sensitive to comparably large scales, but only when $qE$ is reconstructed with high enough signal-to-noise does the FoMr become comparable to what is obtainable with E-mode polarization. The joint FoMr can be quite significant, $\sim 2$, even at the highest CMB noise levels considered, with the inclusion of the remote fields yielding a $\sim 10 \%$ improvement. The joint constraint at the lowest CMB noise levels yields a $\sim 45\%$ constraint on the cutoff scale $k_b$, but the exponent $\delta$ remains very poorly determined.
\end{itemize}

Summarizing, the constraining power on models of large-scale power suppression can in principle be improved by factors $\sim 1.5-2.0$ when E-mode polarization and the remote dipole/quadrupole fields are included. Cosmic variance in measurements of the CMB temperature alone is somewhat mitigated, but due to the very large scales on which there is a significant deviation from a power law spectrum, there are still limitations arising from the finite size of our observable Universe.  

\vspace*{10pt}
\begin{table}[htb]
  \centering
  \renewcommand{\arraystretch}{1.4}
  \resizebox{0.9\textwidth}{!}{
  \begin{tabular}{|c|c|c|c|c|c|}
    \hline
    Model& $\;\;\;$Noise [$\mu$-Karcmin]$\;\;\;$& $\;\;\;$FoMr($T,E$)$\;\;\;$ & $\;\;\;$FoMr($T,v,qE$)$\;\;\;$ & $\;\;\;$FoMr(All)$\;\;\;$ \\
    \cline{1-5}
    \multirow{3}{4cm}{\centering{Exponential suppression}} & 5.0    &      1.37    &    1.37    &   1.61\\
    \cline{2-5}
                                                                                           & 1.0    &       1.37    &    1.53    &    1.70\\
    \cline{2-5}
                                                                                           & 0.1    &       1.37    &    1.68    &    1.83\\
    \cline{1-5}
    \multirow{3}{4cm}{\centering{Broken power law}}          & 5.0    &       1.80    &    1.51    &   2.00\\
    \cline{2-5}
                                                                                           & 1.0    &       1.80    &    1.77    &    2.06\\
    \cline{2-5}
                                                                                           & 0.1    &       1.80    &    2.06    &    2.18\\                                                                      
                                                                                                                                        
    \hline
  \end{tabular}}
\caption{ Figure of merit ratio for the exponential suppression and broken power law models.}
\label{fig:large_sup}
\end{table}

\subsubsection{Features in the power spectrum}

\begin{figure}[htb]
  \centering
    \includegraphics[width=0.6\textwidth]{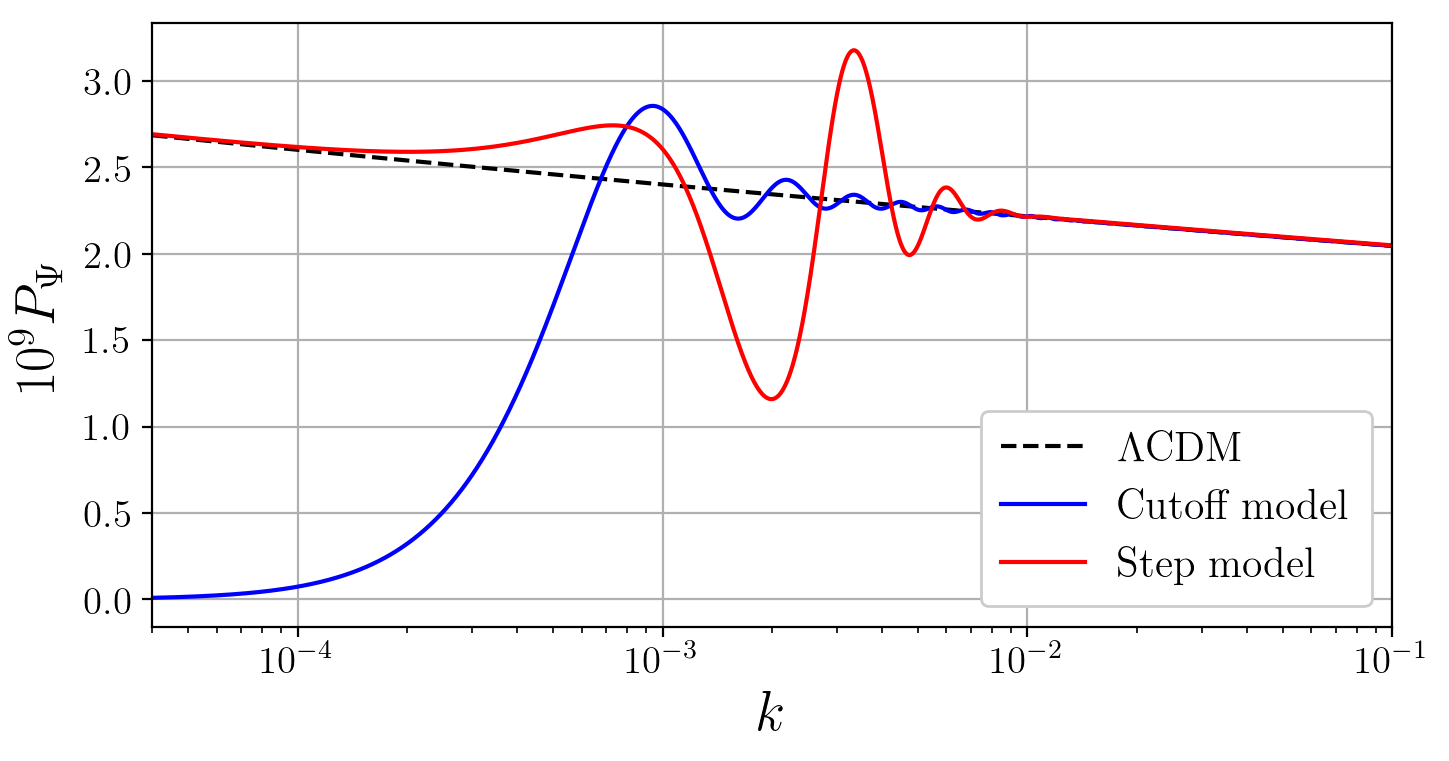}
  \caption{\label{fig:Inflation_models} Primordial power spectrum for the cutoff model and the step model together with the standard $\Lambda$CDM
  spectrum.}
\end{figure}

We now investigate two physical scenarios that give rise to features in the primordial power spectrum. In the first model, we consider a period of kinetic domination preceding slow-roll inflation. This gives rise to a suppression of power on large scales, as well as oscillations in the power spectrum on intermediate scales~\cite{Contaldi_2003}. The one parameter in this model is a scale $k_c$, roughly corresponding to the comoving size of the horizon when slow-roll begins. Clearly, we are able to constrain this model only when there are a minimal number of $e$-folds of inflation, in which case $k_c$ is on observable scales. The full form of the power spectrum is given by:
\begin{equation}
\ln P_{c}(k)=\ln P_0(k)+\ln \left(\frac{\pi}{16} \frac{k}{k_{\mathrm{c}}}\left|C_{\mathrm{c}}-D_{\mathrm{c}}\right|^{2}\right),
\end{equation}
where
\begin{equation}
\begin{array}{l}{C_{\mathrm{c}}=\exp \left(\frac{-i k}{k_{\mathrm{c}}}\right)\left[H_{0}^{(2)}\left(\frac{k}{2 k_{\mathrm{c}}}\right)-\left(\frac{k_{\mathrm{c}}}{k}+i\right) H_{1}^{(2)}\left(\frac{k}{2 k_{\mathrm{c}}}\right)\right]}, \\ {D_{\mathrm{c}}=\exp \left(\frac{i k}{k_{\mathrm{c}}}\right)\left[H_{0}^{(2)}\left(\frac{k}{2 k_{\mathrm{c}}}\right)-\left(\frac{k_{\mathrm{c}}}{k}-i\right) H_{1}^{(2)}\left(\frac{k}{2 k_{\mathrm{c}}}\right)\right]},\end{array}
\end{equation}
and $H_{n}^{(2)}$ denotes the Hankel function of the second kind. We assume the best-fit value from Planck 2015~\cite{Planck_inflation_2015} of $k_c = 3.63 \times 10^{-4} \ {\rm Mpc}^{-1}$ as the central values in our analysis below. The second model we consider arises when there is a tanh-shaped step in the inflaton potential as in Ref.~\cite{PhysRevD.89.083529}, which gives rise to oscillations in the primordial power spectrum. This is a three-parameter model, which, at the level of the inflaton potential, correspond to the location, height, and width of the step. The resulting power spectrum is given by 
\begin{equation}
\ln P_{s}(k)=\exp \left[\ln P_0(k)+I_{0}(k)+\ln \left(1+I_{1}^{2}(k)\right)\right]
\end{equation}
where
\begin{align}
I_{0}(k) &=\frac{\mathcal{A}_{\mathrm{s}} }{2 x^{3}}\left[\left(18 x-6 x^{3}\right) \cos 2 x+\left(15 x^{2}-9\right) \sin 2 x\right]\Big{|}_{x = \left(k / k_{\mathrm{s}}\right)}\times\mathcal{D}\left(\frac{k / k_{\mathrm{s}}}{x_{\mathrm{s}}}\right) \\
I_{1}(k) &=\frac{1}{\sqrt{2}}\bigg{[}\frac{\pi}{2}\left(1-n_{\mathrm{s}}\right)-\frac{\mathcal{A}_{\mathrm{s}} }{x^{3}}\left\{3(x \cos x-\sin x)\left[3 x \cos x+\left(2 x^{2}-3\right) \sin x\right]\right\}\Big{|}_{x = \left(k / k_{\mathrm{s}}\right)}\times\mathcal{D}\left(\frac{k / k_{\mathrm{s}}}{x_{\mathrm{s}}}\right)\bigg{]}.\\
\mathcal{D}(x) &=\frac{x}{\sinh x}\\
\end{align}
We assume the best-fit value from Planck 2015~\cite{Planck_inflation_2015} of $\mathcal{A}_s = 0.374$,  $k_s = 7.94 \times 10^{-4} \ {\rm Mpc}^{-1}$, and $x_s = 1.41$ as the central value in our analysis below. We plot the power spectra for the two models in Fig.~\ref{fig:Inflation_models}. As commented on in Ref.~\cite{Planck_inflation_2015}, for these choices of parameters both of these models give rise to a deficit in the CMB temperature power spectrum at $\ell \sim 20-30$, similar to what is observed.

The results for the FoMr are collected in Table \ref{fig:inflation_models}. More specifically:
\begin{itemize}
\item \bf{Cutoff model}: \normalfont Even at the highest CMB noise levels considered, the remote dipole/quadrupole fields yield an improvement in constraining power comparable to what is achievable with E-mode polarization. The FoMr achievable from all datasets can exceed three, representing a significant improvement in constraining power. Because this model has a combination of power suppression on large scales where the remote quadrupole can be quite informative, and oscillations on scales relevant to the remote dipole field, there is significant improvement on constraints with lowering CMB noise. The Fisher constraints on $k_c$ are quite good, already at the $20\%$-level with CMB temperature alone, and reaching the $6\%$-level when all datasets are included at the lowest CMB noise level considered. However, the posterior distribution for $k_c$ is not Gaussian (indeed, sufficiently small values of $k_c$ should be indistinguishable; see \cite{Contaldi_2003} for more details). This makes interpreting the Fisher constraints, which only probe the curvature of the likelihood function about its maximum, in terms of a statement about model selection impossible. The improved constraining power offered by SZ tomography could yield significant more weight for this model, however a more complete sampling of the likelihood function is necessary to make any definitive conclusion. 

\item \bf{Step model}: \normalfont For the step model, the remote dipole/quadrupole fields prove once again to be competitive with the E-mode polarization. This particular model places oscillatory features on scales at which the remote dipole field is more sensitive than E-mode polarizaion. The reconstructed remote quadrupole adds little information on top of what is offered by the remote dipole. The fully joint analysis for this model yields the highest FoMr considered thus far, reaching nearly a factor of 5 at the lowest CMB noise level considered. Again,  exploration of the full likelihood function is necessary to assess how definitive a detection might be possible with SZ tomography.
\end{itemize}

\vspace*{10pt}
\begin{table}[htb]
  \centering
  \renewcommand{\arraystretch}{1.4}
  \resizebox{0.9\textwidth}{!}{
  \begin{tabular}{|c|c|c|c|c|c|}
    \hline
    Model& $\;\;\;$Noise [$\mu$-Karcmin]$\;\;\;$& $\;\;\;$FoMr($T,E$)$\;\;\;$ & $\;\;\;$FoMr($T,v,qE$)$\;\;\;$ & $\;\;\;$FoMr(All)$\;\;\;$ \\
    \cline{1-5}
    \multirow{3}{4cm}{\centering{Cutoff}} & 5.0    &      2.06    &    2.10    &   2.75\\
    \cline{2-5}
                                                                & 1.0    &       2.06   &    2.56    &    3.06\\
    \cline{2-5}
                                                                & 0.1    &       2.06    &    2.88    &    3.24\\
    \cline{1-5}
    \multirow{3}{4cm}{\centering{Step in inflaton potential}}          & 5.0    &       2.91    &    2.90    &   4.02\\
    \cline{2-5}
                                                                      & 1.0    &       2.91    &    3.45    &    4.46\\
    \cline{2-5}
                                                                      & 0.1    &       2.91    &    3.69    &    4.64\\                                                                      
                                                                                                                                        
    \hline
  \end{tabular}}
\caption{ Figure of merit ratio for the cutoff and step models.}
\label{fig:inflation_models}
\end{table}

\section{Conclusions}\label{conclusions}

Determining whether or not the observed large angular scale anomalies in the CMB are indications of physics beyond $\Lambda$CDM is a matter of great interest and intense debate. Faced with the obstacle imposed by cosmic variance on our study of the largest scales in the Universe, we are driven to analyze data sets that incorporate additional observables on top of the primary CMB temperature in order to favour or rule out the different hypotheses for the origin of the anomalies.

In this paper we explored the constraining power on CMB anomalies models provided by a new set of observables: the remote CMB dipole and quadrupole fields. These fields, which correspond to the projected $\ell =1,2$ moments of the microwave sky as measured at different locations in the Universe, can be reconstructed using SZ tomography. The remote dipole and quadrupole fields carry 3-dimensional information about large scale fluctuations in the Universe, and a significant number of independent modes can in principle be reconstructed from next-generation CMB and galaxy surveys. This additional information is largely independent of the primary CMB, and can therefore offer more statistical power for the analysis of physical models of large scale CMB anomalies.

Our methodology consisted of deriving forecasted constraints on a series of anomalies models using different combinations of observables probing the largest scales, including the primary CMB temperature ($T$), E-model polarization ($E$) and the remote dipole and quadrupole fields ($v,qE$). The improvement on constraints relative to what is achievable with the primary CMB temperature serves as a measure of how informative  additional observables can be, and this was expressed in terms of appropriate figure of merit ratios (FoMr); see Eq.~\ref{FoMr}. We assumed access to data on the full sky, with no systematics (aside from CMB instrumental noise) or foregrounds, in each of our forecasts.

We first considered a model of the power asymmetry where a super-horizon field modulates curvature fluctuations on the largest accessible scales inside our Hubble horizon. We found for this model that the additional information coming from the remote fields can compete and surpass that of the E-mode polarization and that the constraining power as measured by the FoMr can be doubled by including all observables. In principle, this could bring the tension with $\Lambda$CDM to the $5 \sigma$-level. We also investigated constraints on two purely phenomenological models designed to explain the lack of large-scale power in the CMB temperature map. In both cases, the remote dipole/quadrupole add roughly the same additional constraining power as E-mode polarization. The FoMr including all observables reached $\sim 1.5-2.0$, indicating that significant extra constraining power is available with the new observables considered here. Finally, we considered two physical models with features in the primordial power spectrum that could yield the observed power deficit identified in the CMB temperature power spectrum at $\ell \sim 20-30$. The remote dipole field is quite sensitive to the scales on which the primordial power spectrum is modified ($k \sim (10^{-3}-10^{-2})\times\mbox{Mpc}^{-1}$), and the FoMr achievable for these models including all observables can be in the range of $3-4.5$. An improvement at this level could in principle yield a significant detection, although a more systematic investigation of the full likelihood function is necessary to make a definitive and quantitative statement.

Based on the models considered in this paper, we can make a number of general statements about the utility of SZ tomography for addressing the possible physical nature of the CMB anomalies. As many previous analyses have shown~\cite{PhysRevD.77.063008,Copi:2013zja,Yoho:2015bla,ODwyer:2016xov,Bunn:2016kwh,Contreras:2017zjv,Obied:2018qdr,Billi:2019vvg}, E-mode polarization has been identified as a powerful discriminator for physical models of CMB anomalies. For each of the models we have studied, the remote dipole/quadrupole fields offer comparable additional constraining power to E-modes, with fully joint constraints offering even greater improvement. Given the challenges associated with removing foregrounds from E-mode polarization on the largest angular scales, the information provided by SZ tomography could also be highly complementary (although more work is needed to understand the foreground challenges for SZ tomography). Even at the level of sensitivity accessible to current experiments, modulo foregrounds, measurements of E-mode polarization are cosmic variance limited on the largest scales. However, the reconstruction of the remote dipole and quadrupole fields improves significantly with increasingly sensitive CMB experiments. This in turn improves the constraining power for CMB anomaly models. Therefore, in principle, SZ tomography offers a way to systematically improve constraints on CMB anomaly models in the current era of rapidly improving high-resolution, low-noise CMB experiments. In a number of cases, we have shown that SZ tomography could offer the constraining power necessary to determine if anomalies are due to new physics. 

Our analysis has a number of shortcomings. First, our Fisher-based analysis is insensitive to the shape of the likelihood function, which can deviate significantly from a Gaussian for many of the models considered here. A future investigation could improve upon this by sampling the full likelihood function; however given the size of the covariance matrix including all observables and the dimensionality of the parameter space, there will be computational challenges for doing so. Future analyses should also incorporate realistic foregrounds and systematics in the CMB and galaxy surveys, and investigate their impact on the reconstruction of the remote dipole and quadrupole fields. In addition, the effects of masking should be taken into account, which will degrade the information available on the largest angular scales. Despite these limitations, our analysis highlights the new information on the physical nature of the observed CMB anomalies that is {\em in principle} accessible  using SZ tomography. This provides a useful target for future analyses and observations. 

Summarizing, our analysis indicates that there is significant useful information on the large-scale properties of our Universe captured by the reconstructed remote CMB dipole and quadrupole fields using SZ tomography. In combination with the E-mode polarization, which has long been considered one of the main candidates to add statistical power to the study CMB anomalies, we have found that SZ tomography can increase the constraining power on physical models of CMB anomalies by factors of 2-4.5 for achievable levels of CMB instrumental noise. This implies that SZ tomography could play an essential role in determining the physical nature of the CMB anomalies, and motivates further investigation.

\section{Acknowledgments}
We would like to thank Dagoberto Contreras, James Mertens and Moritz M{\"u}nchmeyer for helpful discussions.
This research was supported in part by Perimeter Institute for
Theoretical Physics. Research at Perimeter Institute is supported by the Government of Canada through
the Department of Innovation, Science and Economic Development Canada and by the Province of Ontario
through the Ministry of Research, Innovation and Science. MCJ was supported by the National
Science and Engineering Research Council through a Discovery grant.

\appendix

\section{Constraints on model parameters}\label{sec:models}
We present here the constraints on model parameters derived using Fisher analysis of different combinations of primary CMB and remote dipole and quadrupole fields. We include the constraint using only the primary CMB temperature as well. Characteristic scale parameters $k_c$, $k_b$ and $k_s$ are in units of $\ {\rm Mpc}^{-1}$. \\

\begin{table}[htb]
  \centering
  \renewcommand{\arraystretch}{1.2}
  \begin{tabular}{c} 
  \vspace*{10pt}
\centering{\underline{\small{\textbf{Dipolar modulation model}}}}
  \end {tabular}
  \begin{tabular}{|c|c|c|c|c|c|}
    \hline
    Parameter& $\;\;\;$Noise [$\mu$-Karcmin]$\;\;\;$& $\;\;\;$$\sigma$($T$)$\;\;\;$ & $\;\;\;$$\sigma$($T,E$)$\;\;\;$ & $\;\;\;$$\sigma$($T,v,qE$)$\;\;\;$ & $\;\;\;$$\sigma$(All)$\;\;\;$ \\
    \cline{1-6}
    \multirow{3}{3.5cm}{\centering{$A\:=\:0.07$}}               & 5.0    &    0.021    &    0.015    &    0.014    &   0.012\\
    \cline{2-6}
                                                                                          & 1.0    &    0.021    &    0.015    &    0.012    &    0.010\\
    \cline{2-6}
                                                                                          & 0.1    &    0.021    &    0.015    &    0.011    &    0.010\\
                                                                                                                                                                                            
    \hline
  \end{tabular}
\end{table}

\begin{table}[htb]
  \centering
  \renewcommand{\arraystretch}{1.2}
  \begin{tabular}{c} 
  \vspace*{10pt}
  \centering{\underline{\small{\textbf{Exponential suppression model}}}}
  \end {tabular}
  \begin{tabular}{|c|c|c|c|c|c|}
    \hline
    Parameter& $\;\;\;$Noise [$\mu$-Karcmin]$\;\;\;$& $\;\;\;$$\sigma$($T$)$\;\;\;$ & $\;\;\;$$\sigma$($T,E$)$\;\;\;$ & $\;\;\;$$\sigma$($T,v,qE$)$\;\;\;$ & $\;\;\;$$\sigma$(All)$\;\;\;$ \\
    \cline{1-6}
    \multirow{3}{3.5cm}{\centering{$10^{4}k_c\;=\;3.74$}} & 5.0    &    3.04    &    2.34    &    2.32    &   2.01\\
    \cline{2-6}
                                                                                          & 1.0    &    3.04    &    2.34    &    2.10    &    1.91\\
    \cline{2-6}
                                                                                          & 0.1   &    3.04    &    2.34    &    1.89    &    1.76\\
    \cline{1-6}
    \multirow{3}{3.5cm}{\centering{$\lambda\;=\;0.53$}} & 5.0    &    0.25    &    0.17    &    0.18    &   0.15\\
    \cline{2-6}
                                                                                       & 1.0    &    0.25    &    0.17    &    0.16    &    0.14\\
    \cline{2-6}
                                                                                       & 0.1    &    0.25    &    0.17    &    0.14    &    0.13\\                                                                      
                                                                                                                                        
    \hline
  \end{tabular}
\end{table}

\begin{table}[htb]
  \centering
  \renewcommand{\arraystretch}{1.2}
  \begin{tabular}{c} 
  \vspace*{10pt}
\centering{\underline{\small{\textbf{Broken power law model}}}}
  \end {tabular}
  \begin{tabular}{|c|c|c|c|c|c|}
    \hline
    Parameter& $\;\;\;$Noise [$\mu$-Karcmin]$\;\;\;$& $\;\;\;$$\sigma$($T$)$\;\;\;$ & $\;\;\;$$\sigma$($T,E$)$\;\;\;$ & $\;\;\;$$\sigma$($T,v,qE$)$\;\;\;$ & $\;\;\;$$\sigma$(All)$\;\;\;$ \\
    \cline{1-6}
    \multirow{3}{3.5cm}{\centering{$10^{4}k_b\;=\;5.26$}} & 5.0    &    4.83    &    2.73    &    3.35    &   2.51\\
    \cline{2-6}
                                                                                          & 1.0    &    4.83    &    2.73    &    2.89    &    2.45\\
    \cline{2-6}
                                                                                          & 0.1    &    4.83    &    2.73    &    2.46    &    2.33\\
    \cline{1-6}
    \multirow{3}{3.5cm}{\centering{$\delta\;=\;1.14$}} & 5.0    &    2.80    &    1.52    &    1.77    &   1.35\\
    \cline{2-6}
                                                                                       & 1.0    &    2.80    &    1.52    &    1.49    &    1.30\\
    \cline{2-6}
                                                                                       & 0.1    &    2.80    &    1.52    &    1.30    &    1.22\\                                                                      
                                                                                                                                        
    \hline
  \end{tabular}
\end{table}

\begin{table}[htb]
  \centering
  \renewcommand{\arraystretch}{1.2}
  \begin{tabular}{c} 
  \vspace*{10pt}
  \centering{\underline{\small{\textbf{  Cutoff model  }}}}
  \end {tabular}
  \begin{tabular}{|c|c|c|c|c|c|}
    \hline
    Parameter& $\;\;\;$Noise [$\mu$-Karcmin]$\;\;\;$& $\;\;\;$$\sigma$($T$)$\;\;\;$ & $\;\;\;$$\sigma$($T,E$)$\;\;\;$ & $\;\;\;$$\sigma$($T,v,qE$)$\;\;\;$ & $\;\;\;$$\sigma$(All)$\;\;\;$ \\
    \cline{1-6}
    \multirow{3}{3.5cm}{\centering{$10^{4}k_c\;=\;3.63$}}               & 5.0    &    0.78    &    0.38    &    0.37    &   0.28\\
    \cline{2-6}
                                                                                                       & 1.0    &    0.78    &    0.38    &    0.31    &    0.26\\
    \cline{2-6}
                                                                                                        & 0.1   &    0.78    &    0.38    &    0.27    &    0.24\\
                                                                                                                                                                                            
    \hline
  \end{tabular}
\end{table}

\begin{table}[htb]
  \centering
  \renewcommand{\arraystretch}{1.2}
  \begin{tabular}{c} 
  \vspace*{10pt}
\centering{\underline{\small{\textbf{ $\;\;\;\;$       Step model    $\;\;\;\;$ }}}}
  \end {tabular}
  \begin{tabular}{|c|c|c|c|c|c|}
    \hline
    Parameter& $\;\;\;$Noise [$\mu$-Karcmin]$\;\;\;$& $\;\;\;$$\sigma$($T$)$\;\;\;$ & $\;\;\;$$\sigma$($T,E$)$\;\;\;$ & $\;\;\;$$\sigma$($T,v,qE$)$\;\;\;$ & $\;\;\;$$\sigma$(All)$\;\;\;$ \\
    \cline{1-6}
    \multirow{3}{3.5cm}{\centering{$\mathcal{A}_s\;=\; 0.374$}} & 5.0    &    0.28    &    0.11    &    0.12    &   0.08\\
    \cline{2-6}
                                                                                                   & 1.0    &    0.28    &    0.11    &    0.10    &    0.07\\
    \cline{2-6}
                                                                                                   & 0.1    &    0.28    &    0.11    &    0.09    &    0.07\\
    \cline{1-6}
    \multirow{3}{3.5cm}{\centering{$10^{4}k_s\;=\;7.94$}} & 5.0    &    0.71    &    0.18    &    0.18    &   0.13\\
    \cline{2-6}
                                                                                       & 1.0    &    0.71    &    0.18    &    0.15    &    0.11\\
    \cline{2-6}
                                                                                       & 0.1    &    0.71    &    0.18    &    0.14    &    0.11\\                
                                                                                       
    \cline{1-6}
    \multirow{3}{3.5cm}{\centering{$x_s\;=\;1.41$}}         & 5.0    &    0.60    &    0.25    &    0.23    &   0.18\\
    \cline{2-6}
                                                                                       & 1.0    &    0.60    &    0.25    &    0.20    &    0.16\\
    \cline{2-6}
                                                                                       & 0.1    &    0.60    &    0.25    &    0.18    &    0.15\\                                                                                                                                                      
                                                                                                                                        
    \hline
  \end{tabular}
\end{table}
\FloatBarrier

\section{Mode coupling}\label{sec:modecoupling}

The super-horizon modulating field $h(\vec{\boldsymbol{x}})$ introduced in the spontaneous isotropy breaking mechanism here studied leads to couplings between different multipole moments. The
modified primordial spectrum Eq.~\ref{P_mod} is used to compute the covariance matrix, which differs from the $\Lambda$CDM covariance matrix by terms linear and quadratic in the modulation amplitude $A$:

\begin{align*}
    \begin{split}
        C_{\alpha\beta,\ell\ell',mm'}^{X,Y}  &= C_{\alpha\beta,\ell\ell',mm'}^{X,Y,(\Lambda\mbox{\scriptsize{CDM}})} +  C_{\alpha\beta,\ell\ell',mm'}^{X,Y,(A)} + C_{\alpha\beta,\ell\ell',mm'}^{X,Y,(A^2)}\\[8pt]
        C_{\alpha\beta,\ell\ell',mm'}^{X,Y,(A)} &= \begin{aligned}[t]
        &\delta_{mm'} \sqrt{\frac{4\pi}{3}}\frac{A}{i \chi_{dec}}\int\frac{dk\,k^{2}}{(2\pi)^{3}}P_{\psi}(k)\bigg{[}\Delta^{*X,\alpha}_{\ell}(k)\,\partial_{k}\Delta^{Y,\beta}_{\ell '}(k) - \partial_{k}\Delta^{*X,\alpha}_{\ell}(k)\,\Delta^{Y,\beta}_{\ell '}(k)\\[8pt]
        & - \frac{2\,\Delta^{*X,\alpha}_{\ell}(k)\,\Delta^{Y,\beta}_{\ell '}(k)}{k}\Big{(} \ell\;\delta_{l',l-1}-(\ell+1)\;\delta_{l',l+1}\Big{)} \bigg{]}R^{1\ell'}_{\ell m} \\
        \end{aligned}\\[15pt]
        C_{\alpha\beta,\ell\ell',mm'}^{X,Y,(A^2)}&= \begin{aligned}[t]
        &\delta_{mm'} \frac{4\pi}{3}\frac{A^2}{\chi^{2}_{dec}}\int\frac{dk\,k^{2}}{(2\pi)^{3}}P_{\psi}(k)\sum_{L}R^{1L}_{\ell m}R^{1L}_{\ell' m}
        \bigg{[}\partial_{k}\Delta^{*X,\alpha}_{\ell}(k)\,\partial_{k}\Delta^{*Y,\beta}_{\ell'}(k)\\[8pt]
        &+\frac{\partial_k\Delta^{*X,\alpha}_{\ell}(k)\,\Delta^{Y,\beta}_{\ell '}(k)}{k}\Big{(}(1+\ell')\delta_{L,\ell'-1}-\ell'\delta_{L,\ell'+1}\Big{)}\\[8pt]
        &+\frac{\Delta^{*X,\alpha}_{\ell}(k)\,\partial_k\Delta^{Y,\beta}_{\ell '}(k)}{k}\Big{(}(1+\ell)\delta_{L,\ell-1}-\ell\delta_{L,\ell+1}\Big{)}\\[8pt]
        &+\frac{\Delta^{*X,\alpha}_{\ell}(k)\,\Delta^{Y,\beta}_{\ell '}(k)}{k^2}\Big{(}
        (1+\ell)^2\,\delta_{L,\ell-1}\,\delta_{\ell',\ell}+\ell^2\,\delta_{L,\ell+1}\,\delta_{\ell',\ell}\\
        &-(1+\ell)(\ell-2)\,\delta_{L,\ell-1}\,\delta_{\ell',\ell-2}
        -\ell(\ell+3)\,\delta_{L,\ell+1}\,\delta_{\ell',\ell+2}\Big{)}\bigg{]}\\
        \end{aligned}\\
    \end{split}
\end{align*}

where the couplings $R^{\ell_1\ell_2}_{\ell m}$ are defined through the 3-j Wigner symbols

\be
R^{\ell_1\ell_2}_{\ell m} = (-1)^{m}\sqrt{\frac{(2\ell+1)(2\ell_1+1)(2\ell_2+1)}{4\pi}}\begin{pmatrix} \ell_{1} & \ell_{2} & \ell \\ 0 & 0  & 0 \end{pmatrix}\begin{pmatrix} \ell_{1} & \ell_{2} & \ell \\ 0 & m  & -m \end{pmatrix}.
\ee

The term linear in $A$ induces couplings between multipoles $\ell$ and $\ell\pm1$, while the quadratic term adds couplings between $\ell$ and $\ell\pm2$ as well as the same multipole corrections to the covariance matrix. For the particular case of temperature transfer functions in the Sachs-Wolfe approximation, i.e. $\Delta^{T}_{\ell}(k) \propto j_{\ell}(k\chi_{dec})$, the above expressions can be reduced to 
those presented in \cite{PhysRevD.72.103002,PhysRevD.77.063008} using appropriate recursion relations for the derivatives of the spherical Bessel functions. A similar approach to compute
the multipole couplings in terms of derivatives of the transfer functions was taken in reference~\cite{Mnras/stw1039}, although the assumptions on the modulation of the primordial spectrum are not the same
as ours and the $\mathcal{O}(A^2)$ term was not computed.

\bibliography{references}

\end{document}